# Far-From-Equilibrium Physics:  An Overview

**Heinrich Jaeger, James Franck Institute and Department of Physics, University of Chicago**
**Andrea J. Liu, Department of Physics and Astronomy, University of Pennsylvania**

*Isolated systems tend to evolve towards equilibrium, a special state that has been the focus of many-body research for a century.  Yet much of the richness of the world around us arises from conditions far from equilibrium.  Phenomena such as turbulence, earthquakes, fracture, and life itself occur only far from equilibrium. Subjecting materials to conditions far from equilibrium leads to otherwise unattainable properties.  For example, rapid cooling is a key process in manufacturing the strongest metallic alloys and toughest plastics.  Processes that occur far from equilibrium also create some of the most intricate structures known, from snowflakes to the highly organized structures of life. While much is understood about systems at or near equilibrium, we are just beginning to uncover the basic principles governing systems far from equilibrium.*

## THE IMPORTANCE OF FAR-FROM-EQUILIBRIUM PHENOMENA

We live in a world of evolving structures and patterns. When energy is continually supplied to systems with many interacting constituents, the outcome generally differs strikingly from the unchanging state that characterizes equilibrium. From the molecular processes on the nanoscale that form the basis of life to the dynamically changing climate on this planet to the clustering of matter within the universe as a whole, a myriad of phenomena owe their existence to being not just slightly away from equilibrium, but far from it (Fig. 1).  Far-from-equilibrium conditions also strikingly alter the behavior of ordinary fluids and solids.  Dramatic examples occur when fluid flow turns turbulent or when solids give way and fracture (Fig. 2).  Both turbulence (Falkovich *et al.* 2001) and fracture (Carlson *et al.* 1994; Freund 1998) generate patterns of amazing complexity that not only completely change the materials properties but also redistribute energy across a whole hierarchy of nested structures ranging from the microscopic to the macroscopic scale.  Far-from-equilibrium processes span a similarly immense range of time scales, from electronic transitions at the sub-nanosecond scale, to glassy relaxation too slow to measure with any technique, to the age of the universe.

Far-from-equilibrium behavior is not confined to special conditions or certain types of materials. Instead, it arises across the entire spectrum of physics in a host of problems of fundamental interest.  Far-from-equilibrium phenomena also benefit and plague us in technology and in everyday life (Ennis *et al.* 1994; Marder and Fineberg 1996) (Fig. 2).

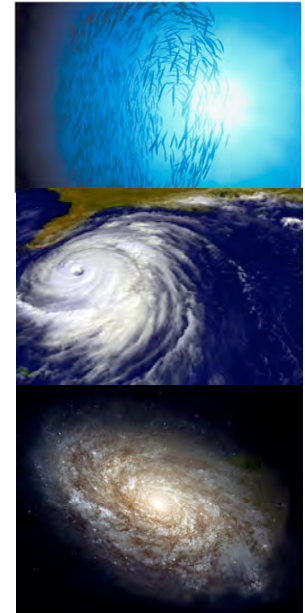

Fig.1.  Swarming schools of fish, swirling storms and galaxies (top to bottom) are all examples of systems formed and evolving far from equilibrium.

Indeed, some of the most complex outcomes of behavior far from equilibrium emerge in situations we are familiar with from daily experience. We can see turbulence in cloud patterns as well as in a bath tub; we take advantage of glassy behavior in nearly all plastics but suffer from it in traffic jams; we exploit the breaking-up of a stream of fluid into droplets with fuel injection and ink jet printing but also find it in every leaky faucet. The reach of far-from-equilibrium phenomena extends even further to many systems of profound societal importance. In the last decade, researchers from the condensed-matter physics community have begun to tackle far-from-equilibrium behavior governing the workings of systems ranging from the economy to ecosystems and the environment (Wu and Loucks 1995; Ghashghaie *et al.* 1996; Mantegna and Stanley 2000; Sornette 2003; McCauley 2004; Moritz *et al.* 2005).

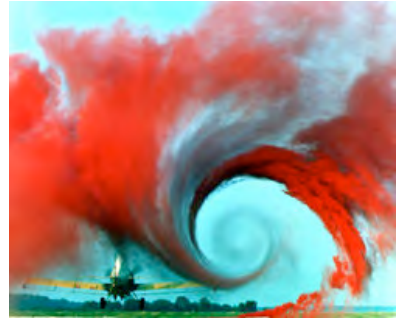

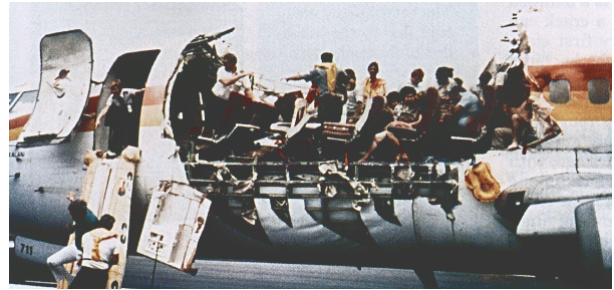

Fig.2. The need to control far-from equilibrium behavior. Top: Turbulent airflow produced by the wingtips of a small airplane (visualized by red smoke). Bottom: Disastrous effect of materials fatigue and eventual fracture.

Two important themes define the scope of the challenge, and they run as persistent motifs through a description of the current status of physics far from equilibrium. First, ***far-from-equilibrium behavior is ubiquitous***. The breadth of phenomena investigated makes the study of far-from-equilibrium systems an inherently interdisciplinary field that forges connections between the physics community and researchers in biology, chemistry, applied mathematics, geology, meteorology and engineering. Far-from-equilibrium physics is connected intimately to both fundamental scientific challenges and cutting-edge materials processing. Finally, far-from-equilibrium physics underlies a wide range of phenomena outside the traditional boundaries of condensed matter physics, including earthquakes, hurricanes, galaxy formation, and consciousness. As a result, breakthroughs in the area have potential for far-reaching impact.

The second key theme is that far-from-equilibrium behavior is not a simple extension of equilibrium or near-equilibrium physics. Instead, it corresponds to qualitatively different types of behavior and response, typically associated with crossing some threshold into a new regime. In some specific cases we have been able to unearth the microscopic origins of far-from-equilibrium phenomena, but we are still lacking the understanding necessary to develop more comprehensive frameworks. Therefore, despite its importance, ***far-from-equilibrium behavior still remains largely uncharted territory***. The reasons why far-from-equilibrium phenomena often resist understanding are described below. The is followed by a discussion of problems for which we have been able to identify robust features, both in experiment and in theory, that can serve as starting points for future work. Except for small modifications and for the addition of references, this article corresponds to Chapter 5 of the CMMP2010 report released by the National Research Council (NRC *et al.* 2007). In order to provide entering researchers a starting point, we focused primarily on reviews, when available, in selecting the references. Clearly this glosses over much of the relevant literature. For a more exhaustive treatment we ask the reader to turn to the more specialized reviews in the references.

**What Condensed Matter Physicists Bring to the Table**

Researchers trained in condensed matter physics are well-positioned to spearhead progress in the field of far-from-equilibrium behavior. As one of the forefront areas of interdisciplinary research, condensed matter physics has long been a focal point for new approaches that bring together ideas from physics and other science and engineering disciplines, and that connect basic science with applied research. Condensed matter physics also specializes in developing new theoretical, numerical, and experimental tools and techniques for systems of many interacting constituents. Experimental techniques that have been especially useful for probing far-from-equilibrium behavior include novel imaging tools, spectroscopic and particle tracking methods, and optical tweezers. Many powerful theoretical and numerical techniques for studying the emergent behavior of many-particle systems near equilibrium have been generalized to systems far from equilibrium; for example, techniques originally developed for studying magnets have been extended to the flocking of birds (Toner *et al.* 2005). Perhaps the field's most valuable characteristic, however, is its penchant for searching for commonalities in wildly disparate systems. This focus led to the spectacular success of condensed matter physics in realizing that the enormous variety of equilibrium phase transitions can be understood in terms of a few classes of behavior. This history motivates researchers to search for similar organizing principles in the even vaster array of far-from-equilibrium phenomena.

Far-from-equilibrium behavior is an important component in several of the challenges discussed in the CMMP2010 report (NRC *et al.* 2007). It underlies many emergent phenomena in systems ranging from the nanoscale to the macroscale, and it plays an essential role in the physics of living systems. Because many far-from-equilibrium phenomena require energy in order to be driven, they are also inevitably implicated in energy consumption and conversion. In quantum computing, the challenge is to prepare qubits in prescribed pure quantum states. Such systems are necessarily far from equilibrium. Finally, because far-from-equilibrium phenomena are so common in everyday life and underlie so many societal concerns, they provide a rich context for education and learning, for the next generation of scientists as well as for the general public.

# HOW DO SYSTEMS REACH THE FAR-FROM-EQUILIBRIUM REGIME AND WHAT MAKES FAR-FROM-EQUILIBRIUM PHYSICS DIFFICULT?

One way to keep a system from its natural state of rest and push it into the far-from-equilibrium regime is to subject it to continual and sufficiently strong forcing. For example, the energy that continually strikes the earth from the sun gives rise to far-from-equilibrium behavior ranging from chaotic weather patterns to the staggering diversity of life. If solar energy were no longer supplied, many systems on earth would revert to equilibrium. Driven systems such as these not only give rise to rich and unanticipated phenomena but are also of tremendous importance to technological applications. For example, in molecular or nano-scale electronics, new phenomena arise from the response to large electromagnetic fields, currents and mechanical stresses (Gaspard 2006). As one scales the physical dimensions of matter to the nanometer scale, the applied fields that drive the system away from its equilibrium state are amplified while the scattering that allows relaxation back to equilibrium is suppressed (Huang *et al.* 2007). As a result, such devices often operate in the far-from-equilibrium regime, unlike conventional semiconductor devices at the micron scale, which typically operate much closer to equilibrium. On somewhat larger scales, far-from-equilibrium processes such as grinding and milling have been key to many industries for a long time, and newer techniques such as mechanical alloying of powders are gaining attention for novel materials synthesis (Murty and Ranganathan 1998; Suryanarayana 2001).

Conditions far from equilibrium also provide a route for controlling a larger variety of patterns and for assembling structures from the nanoscale on up at growth rates much faster than would be possible with near-equilibrium approaches (Stoldt *et al.* 1998; Lopes and Jaeger 2001; Rabani *et al.* 2003). Importantly, far-from-equilibrium processes can achieve structural and dynamical richness even with the simplest of ingredients, such as the intricate dendritic growth realized in snowflakes (Langer 1980) (Fig.3).

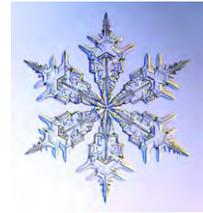

Fig.3. Far-from-equilibrium growth in nature: the snowflake

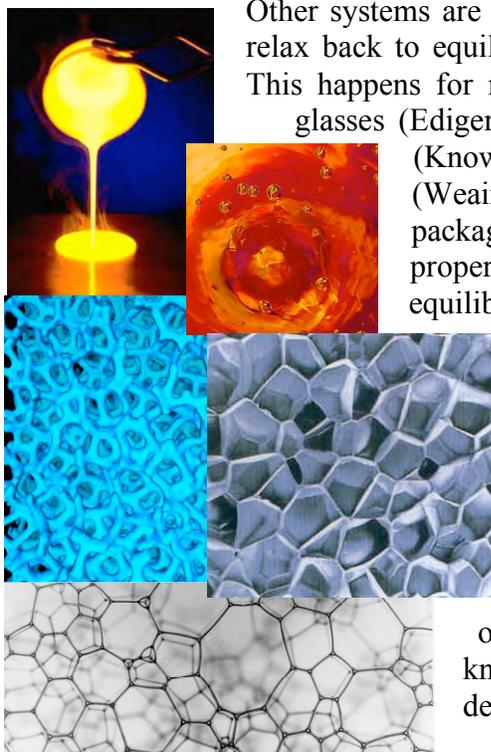

Fig.4. Glasses and foams are examples of important materials that are generically in states far from equilibrium. Clockwise from top: molten glass freezing into a solid, Styrofoam, soap foam, high-strength light-weight nickel foam.

Other systems are trapped far from equilibrium because they simply cannot relax back to equilibrium even after all driving forces have been removed. This happens for many materials vital to our industrial society, including glasses (Ediger *et al.* 1996; Debenedetti and Stillinger 2001), powders (Knowlton *et al.* 1994; Shinbrot and Muzzio 2000), foams (Weaire and Hutzler 1999; Banhart 2001) and polymeric packaging materials (Auras *et al.* 2004), which attain their properties from being intrinsically caught in far-from-equilibrium states. These materials exhibit structural properties that, under equilibrium conditions, would identify them as liquids; yet they can behave like solids (Fig. 4).

Processes that occur far from equilibrium are beginning to force us to rethink some of the foundations of condensed matter and materials physics. Yet even today, most of our knowledge about how systems with many constituent particles behave and evolve is based on considerations valid only close to equilibrium. We know much more about systems near equilibrium and have developed a powerful formalism, statistical mechanics, to predict the emergent, collective behavior of many-particle systems. This framework has allowed condensed matter researchers to understand a large number of phases of matter, the origins of many of their properties and the nature of transitions between them. However, this framework applies only to situations where a system is thermally and mechanically in balance with its surroundings, and thus it covers only a small subset of the phenomena we observe around us and that we confront in applications.

One conceptual difficulty posed by systems far from equilibrium thus arises from the absence of established theoretical frameworks. By virtue of being far from equilibrium, such systems pose numerous challenges. They are typically *nonlinear*; *i.e.* their response to perturbation is often not proportional to the magnitude of the perturbation, as for systems near equilibrium. They are often *disordered*; *i.e.* their structure is typically not crystalline, as equilibrium solids generally are. Finally, they are often *non-ergodic*; *i.e.* they do not necessarily explore a large subset of the states available to them, as equilibrium systems must. As a result, even characterizing their behavior and structure leads one onto largely unfamiliar ground from the standpoint of most of condensed matter physics.

### Far-From-Equilibrium Materials

Certain classes of materials almost always exist under conditions far from equilibrium. Many materials investigated by soft condensed matter researchers fall into this category, including glasses, foams, granular materials and dense colloidal suspensions (Liu and Nagel 2001). In all these examples, the thermal energy supplied by the surroundings is too small to allow the systems explore many configurations. Instead, they are trapped in configurations that structurally resemble a liquid (they are dense and highly disordered), but are unable to flow and thus behave as solids. This glassy behavior, a hallmark of many far-from-equilibrium materials, is observed for constituents ranging from molecules in glass-forming liquids to grains of sand in dunes. Since these materials cannot relax to equilibrium, they typically retain a memory of the preparation or processing conditions, a key for many technological innovations such as molded plastic parts and shape memory polymers (Gunes and Jana 2008). Transitions from far-from-equilibrium glassy states to near-equilibrium crystalline states are the basis for chalcogenide glass optical disks (Zakery and Elliott 2003) and phase-change memory devices (Wuttig and Yamada 2007).

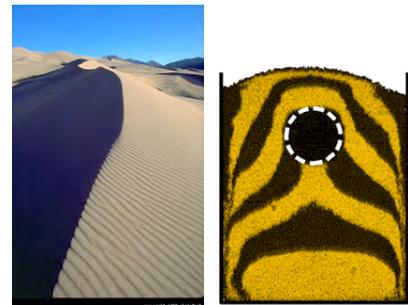

Fig.5. Granular materials consist of individually solid grains, interacting only at contact; yet large assemblies of such grains exhibit a rich set of complex behaviors. Left: ripples in a sand dune. Right: while fluids mix when stirred, granular materials size-separate. MRI of interior of layered granular system, showing upward motion of large particle (dotted circle) in a bed of smaller ones.

Over the last decade, granular matter has emerged as a key prototype of a far-from-equilibrium material (Jaeger *et al.* 1996; de Gennes 1999; Kadanoff 1999) (Fig. 5). In its simplest form, granular matter consists of nothing more than a large number of non-cohesive, macroscopic hard spheres interacting only at contact; yet it exhibits all the characteristics of far-from-equilibrium behavior, as discussed in the subsequent sections. Furthermore, several ideas developed originally within the context of granular materials have by now been successfully "exported" into other areas; for example, the concept of jamming gives insight into glassy phenomena (Liu and Nagel 2001). Similarly, ideas about avalanche statistics in driven dissipative systems, investigated early on in sand piles, have been applied to the dynamics of traffic (Helbing 2001), earthquakes (Carlson *et al.* 1994) and friction (Nasuno *et al.* 1997; Bretz *et al.* 2006), and have been connected to magnetic-flux-bundle motion in superconductors (Jaeger and Nagel 1992; Field *et al.* 1995; Nowak *et al.* 1997; Olson *et al.* 1997; Altshuler and Johansen 2004; Aranson *et al.* 2005). Beyond fundamental research, a large number of industrial processes depend on the handling and transport of granular matter, from seeds and fertilizer pellets in agriculture to ore and gravel in mining operations to powders and pills in the pharmaceutical industry (Ennis *et al.* 1994; Knowlton *et al.* 1994; Shinbrot and Muzzio 2000). Yet the inherently far-from-equilibrium behavior of these materials is still poorly understood and controlled. For example, in North America, new plants designed for processing granular materials initially operate at only about 50-60% of design capacity, while those designed for the handling of liquids immediately operate at nearly full efficiency (Merrow 1985; Merrow 1988).

**Far-From-Equilibrium Processing and Assembly**

Many materials processing techniques exploit far-from-equilibrium conditions for the growth and manufacture of materials that otherwise could not be fabricated. High-strength alloys are often formed by the same rapid dendritic growth that underlies the formation of snowflakes. Some of the very strongest materials available are metallic alloy glasses, made by rapid cooling into amorphous states far from equilibrium (Inoue 2000; Wang *et al.* 2004) (Fig. 6). Lightweight, strong and tough plastics for car bumpers and aircraft are produced by similar processes. The understanding and control of out-of-equilibrium behavior are also important for interface growth processes, for example during the mechanical alloying already mentioned above (Murty and Ranganathan 1998; Suryanarayana 2001) or the fabrication of large, defect-free crystalline domains for silicon thin film devices by laser-induced melting and solidification (Sposili and Im 1996; Hatano *et al.* 2000).

Far-from-equilibrium processing conditions can be used to drive a system towards unique final configurations in very efficient and speedy ways. On the nanoscale this offers new advantages. For example, certain polymers (diblock-copolymers) spontaneously organize themselves into extended patterns with repeat spacings in the 10-50nm range (Park *et al.* 2003; Hawker and Russell 2005). Such feature sizes are difficult to achieve with conventional lithographic methods and are desirable for applications ranging from drug delivery (Allen *et al.* 1999) to high-density magnetic storage (Park *et al.* 1997; Ross *et al.* 1999; Kim *et al.* 2003; Terris and Thomson 2005). Unprocessed, these polymeric structures are often fairly disordered and contain a large number of defects. If the systems are sheared, however, the defects can be removed and the structures can order over extremely large distances (Chen *et al.* 1997; Angelescu *et al.* 2004).

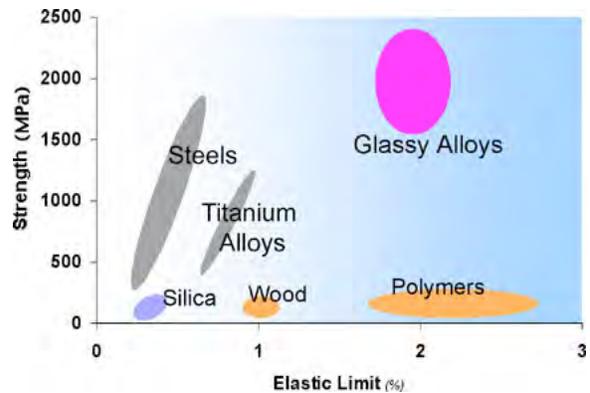

Fig.6. Far-from-equilibrium processing produces some of the highest strength materials (glassy metal alloys).

An advantage of far-from-equilibrium processing conditions is that small differences in the physical or chemical properties of neighboring regions in a material can be amplified; in equilibrium, diffusion tends to smooth out such differences. This property can be exploited to aggregate inorganic components, such as metallic, magnetic or semiconducting particles, on selective polymer domains, leading to new nanoscale-structured hybrid materials (Lopes and Jaeger 2001; Thompson *et al.* 2001). Another important area where far-from-equilibrium conditions can be used to control morphologies is drying-mediated assembly of nanoparticles. Here pathways have been developed for the creation of both densely packed but disordered thin film aggregates (Rabani *et al.* 2003; Sztrum and Rabani 2006; Lee *et al.* 2007) and highly ordered superlattices (Bigioni *et al.* 2006; Xu *et al.* 2007).

## WHAT DETERMINES THE BEHAVIOR FAR-FROM-EQUILIBRIUM?

In equilibrium, minimization of a free energy determines the preferred state, and the system reaches this state independent of the initial conditions. Far from equilibrium, systems typically exhibit a very rich set of characteristic behaviors that are not generally described by a minimization principle. What physics governs the state a system chooses? We have made considerable progress in a number of specific cases. This section discusses advances in the areas of fluids and dynamical systems, and looks at the use of singularities in understanding and controlling far-from-equilibrium behavior.

### Systems with Hydrodynamic Equations of Motion

In many cases, far-from-equilibrium systems exhibit a convenient separation of length and time scales. In order to understand many fluid-flow problems, such as the vortex of a tornado, it is not necessary to describe the motions of individual molecules. Experience with systems at or near equilibrium teaches us that it is often fruitful to focus on the long length scale, long time scale behavior (Chaikin and Lubensky 1995). This so-called "hydrodynamic" approach has been the basis of success in describing a number of systems far from equilibrium. Once we know the basic differential equations that describe long length and time scale properties we can tackle an astounding range of far-from-equilibrium, nonlinear behaviors. The prototypical example is the Navier-Stokes equation for incompressible fluid flow:

$$\rho\left(\frac{\partial \vec{v}}{\partial t} + \vec{v} \cdot \nabla \vec{v}\right) = -\nabla p + \eta \nabla^2 \vec{v} \qquad (1)$$

Like other typical hydrodynamic equations of motion, the Navier-Stokes equation is a nonlinear partial differential equation that describes long length and time scale properties and can be used to tackle an astounding range of far-from-equilibrium, nonlinear behaviors, including erratic fluttering of flags in the wind (Zhang *et al.* 2000) (Fig. 7), the flapping of a bird's wing (Vandenberghe *et al.* 2004), and the breaking of water waves on a beach (Lin and Liu 1998). The difficulty lies in solving the equation. Similar descriptions also apply to complex fluids under flow, a frontier area that is only beginning to be explored (Morozov and van Saarloos 2007; White and Mungal 2008). Finally, the hydrodynamic approach can be applied to a wide range of phenomena not immediately associated with fluids, such as collapsing white dwarves (Plewa *et al.* 2004; Fisker *et al.* 2006), the flocking of birds and other organisms (Toner *et al.* 2005), and the development of single-celled amoebae into multicellular organisms (Ben-Jacob *et al.* 2000). Another example is found in semiconductor heterostructures in which electron density waves are confined to the sample edge. There, when dissipation plays no role, strong electronic correlations are predicted (Bettelheim *et al.* 2006) to produce dispersive shock waves that resemble roll clouds in the atmosphere (Christie 1989). Many more examples are provided by the physical, chemical and biological systems that exhibit pattern formation, in which a uniform system develops patterns in space and/or time by being driven out of equilibrium (Cross and Hohenberg 1993).

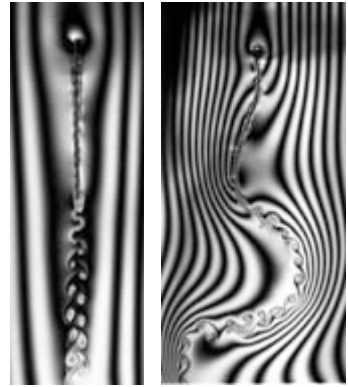

Fig.7. The transition from steady (left) to fluttering (right) motion, visualized by imaging the fluid flow around a filament tied to a post (circle at the top).

The idea that the dynamics of a system with many degrees of freedom can be dominated by the interaction of only a few (such as those at long length and time scales) is an important concept that motivates the study of simple dynamical models to gain insight into complex phenomena. For example, the Lorenz model is a set of nonlinear coupled equations for three variables:

$$\frac{dx}{dt} = \sigma\left(y - x\right)$$
$$\frac{dy}{dt} = x\left(\rho - z\right) - y \qquad (2)$$
$$\frac{dz}{dt} = xy - \beta z$$

where $\sigma$, $\rho$ and $\beta > 0$ are adjustable parameters. The Lorenz model was originally developed to describe convection rolls in the atmosphere and includes only a few degrees of freedom yet successfully capture qualitatively many features of the earth's climate (Palmer 1993). Similar approaches are used to gain insight into the origin of the earth's magnetic field (Roberts and Glatzmaier 2000), mantle convection (Ogawa 2008), Jupiter's red spot (Marcus 1993), and the cycle of solar flares (Low 2001).

The challenges in tackling this class of problems lie in the identification of the few crucial degrees of freedom that must be retained and in the complexity of the resulting equations of motion. Much progress has come from a close coupling of analytic theory with large-scale computer simulations, informed by experiments.

## Turbulence and fracture

Many far-from-equilibrium phenomena pose special challenges because they involve a multitude of length and time scales that interact and thus all become important. Large-scale turbulence is connected directly to flow behavior at scales many orders of magnitude smaller (Mellor and Yamada 1974; Falkovich *et al.* 2001); macroscopic fracture patterns depend intimately on the local configuration of molecular bonds in front of the crack tip (Marder and Fineberg 1996; Freund 1998). In problems such as turbulence, hydrodynamic equations apply but become impossible to solve. Theoretical techniques used in condensed matter physics to study equilibrium critical phase transitions, such as the renormalization group, can be useful here. These techniques are designed to understand how physics at small length or time scales affects behavior at somewhat larger length or time scales, and so on, ultimately leading to an understanding of how a wide range of length scales or time scales interact with one another.

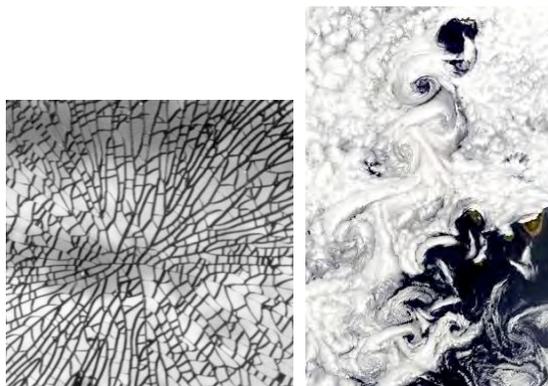

Fig. 8. Far-from-equilibrium behavior often involves processes interacting over a large range of length (and time) scales, leading to characteristic patterns such as the ones observed in the fracture of a glass (left) and in turbulent cloud formations (right; over the Canary Islands).

We will focus here on turbulence as one of the most common far-from-equilibrium phenomena in the environment and in industrial processes. Turbulence produces complex flow structures that modify the transport of momentum, mass and heat, thereby creating a wide variety of both wanted and unwanted effects: a means for rapid mixing of reagents in industrial processes but also parasitic drag in pipe flow and, on a larger scale, catastrophic weather patterns such as hurricanes. Very similar unstable flow structures are produced during the extrusion of polymers or pastes through an orifice (Meulenbroek *et al.* 2003), in flows of complex fluids such as polymer solutions ("elastic turbulence") (Groisman and Steinberg 2000), and in slow sedimentation of particles at high concentration in a fluid (Segre *et al.* 2001). Thus, the mechanisms underlying turbulence appear to be remarkably general. Ideas from turbulence have even been applied to finance (Ghashghaie *et al.* 1996; Mantegna and Stanley 2000). Despite the ubiquity and importance of turbulence, however, we do not understand how it develops well enough to control or prevent it in many cases. The onset flow rate and the nature of the onset of turbulence are still puzzling; turbulence often sets in gradually, in stages, but in many cases, including simple pipe flows, turbulence sets in prematurely for reasons that remain vexingly elusive (Hof *et al.* 2003; Faisst and Eckhardt 2004). Finally, despite much progress during the last decade or two, the nature of the fully-turbulent state still poses many open problems (Siggia 1994; Kadanoff *et al.* 1995; Lewis and Swinney 1999; Sreenivasan 1999; La Porta *et al.* 2001; Tabeling 2002; Benzi *et al.* 2008). In this state, long-lived, large-length-scale coherent structures play an important but still poorly understood role. In the next decade, particle-tracking techniques (Ott and Mann 2000; Ouellette *et al.* 2006) for imaging fluid elements during turbulent flow should shed light on many of these longstanding questions.

## Singularities

In many circumstances, especially under extreme mechanical loading or shearing conditions, materials are driven so far from equilibrium that they change their shape irreversibly. This happens every time a liquid splashes and breaks up into droplets, a piece of glass fractures, a sheet of paper crumples, or a car crashes. Such catastrophic events are typically connected with deformations or failure modes that act at the smallest possible scales and yet affect the overall shape. Consider a slowly dripping faucet with water that is just about to pinch off into a drop. What sets the shape of drop and of the neck by which it hangs just before breaking off? It turns out that, in many cases, these shapes are controlled completely and at every stage by only one spot along the neck, namely where the neck is thinnest (Cohen *et al.* 1999; Eggers and Villermaux 2008). This type of behavior is *scale invariant*—an image of a neck gives no clue as to the overall size of the neck. In other words, the breaking-apart into a drop is controlled by a local singularity, in this case the divergence of the neck curvature. One important recent realization has been that there are two classes of singular break-up events: one in which the neck shape is universal since it does not remember the initial conditions, and one where there is memory of the early stages of neck formation and thus no universality. An everyday example of the latter one is an air bubble breaking off and then rising inside a highly viscous medium such as honey (Doshi *et al.* 2003; Keim *et al.* 2006). Similarly, the overall behavior of a crumpled piece of paper is determined by small number of local spots, sharp points of very high curvature connected by a network of ridges (DiDonna and Witten 2001; Witten 2007). Such singular spots instantly transform an otherwise floppy sheet into a structure that can bear loads and absorb shocks (Fig. 9). Similar stress focusing has also been observed at the nanoscale, in crumpled sheets of nanoparticles (Lin *et al.* 2003).

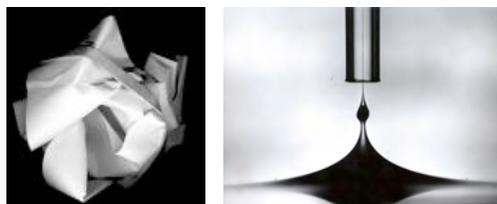

Fig.9. Using singularities to control materials' properties. Left: crumpling a piece of paper stiffens it and allows it to absorb shock. Right: entraining particles, for encapsulation purposes, into the near-singular flow when an interface between two fluids (here: oil and water) is deformed by extruding the oil with a pipette.

Similar scale invariance occurs at singularities such as those at critical phase transitions in equilibrium systems. Over the last decade, researchers have built upon the conceptual foundation of equilibrium phase transitions to identify and tackle far-from-equilibrium materials under extreme conditions. These systems were previously intractable precisely because of their singularities; the triumph in the last decade has been to *exploit* singularities to understand how they control the behavior of such systems over a broad range. Extensions of this approach have demonstrated how the unique behavior in the vicinity of a singularity can be used to achieve unprecedented levels of processing control, which can be used, for example, to uniformly encapsulate live cells prior to transplantation (Fig. 9, right) (Cohen *et al.* 2001; Wyman *et al.* 2007). The extreme mechanics associated with singularities are likely to become increasingly important. They also are prime examples of how, far from equilibrium, the evolution of structure and dynamics are often inseparable.

**Robustness as a Design Principle**

In the last decade, ideas from engineering and biology have led physicists to explore a mechanism of state selection very different from equilibrium free energy minimization. Many far-from-equilibrium systems have been designed, either by deliberate engineering or through evolution and natural selection, to be robust to perturbations (Barkai and Leibler 1997; Carlson and Doyle 2002; Gao *et al.* 2003). For example, cars are now designed with complicated internal networks involving many components, backup mechanisms and adaptive feedback loops to ensure reliable operation under a wide range of environmental conditions. Likewise, biological networks, such as those that enable white blood cells to pursue invading bacteria, have evolved to be insensitive to biochemical changes in their components (Barkai and Leibler 1997; Alon *et al.* 1999). Recently, researchers have used maximization of robustness as a mechanism of state selection in interacting networks (Albert *et al.* 2000; Strogatz 2001; Albert and Barabasi 2002; Milo *et al.* 2002; Newman 2003) (Fig. 10). This opens up a vast array of systems that can be studied using the tools of condensed matter physics, ranging from circadian clocks (Tyson *et al.* 2008)to the Internet (Albert and Barabasi 2002; Newman 2003)and from our immune systems (Jerne 1974) to the environment (Proulx *et al.* 2005). One interesting common feature of systems designed for robustness is that their complexity renders them vulnerable to rare, unexpected perturbations (Carlson and Doyle 2002). For example, the network of interconnected species in the world's oceans has adapted over millennia to be remarkably stable to the vast number of perturbations that can occur. Yet a small change of acidity in ocean waters produced by increased carbon dioxide in the atmosphere may trigger mass extinctions of species (Orr *et al.* 2005; Guinotte and Fabry 2008). Even far-from-equilibrium systems, such as materials under stress, which have not evolved or been specifically designed, can exhibit similar vulnerabilities, such as fracture, due to the history of their formation and the complexity of interactions among the many atoms or molecules that constitute them.

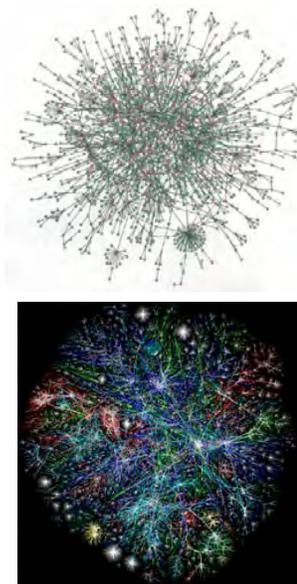

Fig. 10. Examples of evolving network structures far from equilibrium. Top: map of interacting yeast proteins. Bottom: Internet nodes.

**Predictability and Control: What can we learn from fluctuations?**

For systems comprised of many particles in or near equilibrium, statistical mechanics tells us that fluctuations of observable quantities around their average values tend to be small and to have a Gaussian (bell-shaped) distribution. For systems far from equilibrium, there is no general framework such as statistical mechanics, and fluctuations tend to be distributed rather differently. The distributions are often broader than simple Gaussian "bell curves", exhibiting for example slower, exponential decays, power-laws or additional peaks, so that catastrophic, but rare, events can dominate behavior. This is the case in avalanches involving sudden magnetic domain reorientations or flux bundle motion in superconducting magnets (Jaeger and Nagel 1992; Field *et al.* 1995; Olson *et al.* 1997; Aranson *et al.* 2005) (Fig. 11). Similar fluctuations that include large, catastrophic events occur in granular materials, as in landslides or mudslides, or during earthquakes (Carlson *et al.* 1994; Jaeger *et al.* 1996; Bretz *et al.* 2006). Turbulence and spatio-temporal chaos also produce characteristic fluctuations in the measured quantities (Lewis and Swinney 1999; Falkovich *et al.* 2001; La Porta *et al.* 2001; Ouellette *et al.* 2006; Benzi *et al.* 2008). In sheared granular materials, where stresses propagate along highly branched networks of "force chains", the characteristic distribution of inter-particle contact-force magnitudes around their mean can give valuable information about incipient failure, the transition from jammed to unjammed behavior, and the nature of the flowing state (Liu *et al.* 1995; Howell *et al.* 1999; Brujic *et al.* 2003; Corwin *et al.* 2005; Majmudar and Behringer 2005; Behringer *et al.* 2008). Quite generally, the spectrum of fluctuations thus can serve as a key signature of far-from-equilibrium behavior and a powerful tool to analyze the underlying mechanism.

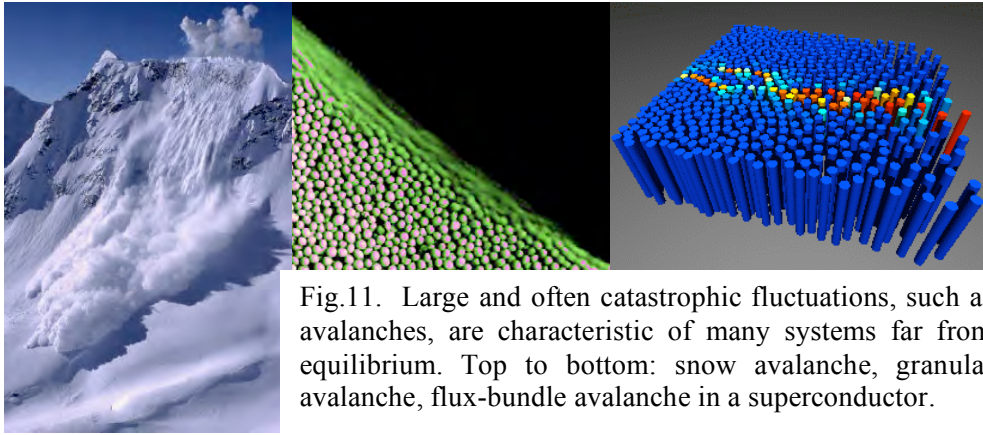

Fig.11. Large and often catastrophic fluctuations, such as avalanches, are characteristic of many systems far from equilibrium. Top to bottom: snow avalanche, granular avalanche, flux-bundle avalanche in a superconductor.

One of the most important questions one can ask about a many-particle system is how it will respond to perturbations. For systems in thermal equilibrium, the fluctuation-dissipation theorem provides the answer: we know that if the perturbation is small, the system will respond just as it does to naturally occurring fluctuations. The relationship between correlation and response depends on temperature;. For example, the second law of thermodynamics states that fluctuations of any extensive thermodynamic variable $X$ are related to the response of the average value of $X$ to its thermodynamic conjugate $\xi$ by:

$$\left\langle \left( X - \langle X \rangle \right)^2 \right\rangle = -k_B T \frac{\partial \langle X \rangle}{\partial \xi} \qquad (3)$$

Thus, temperature measures the size of fluctuations relative to the response, which quantifies how hard it is to create a fluctuation. In other words, one could define temperature by

$$k_B T = \frac{\left\langle \left( X - \langle X \rangle \right)^2 \right\rangle}{-\partial \langle X \rangle / \partial \xi} \qquad (4)$$

For systems far from equilibrium temperature no longer plays such a role. However, in analogy to the thermal case it is possible, in some cases, to define an *effective* temperature from the relationship between correlation and response. For certain classes of driven dissipative systems, such as sheared glasses, foams, or fluidized granular materials such as vibrated or gas-fluidized granular beds, there is evidence that the notion of an effective temperature can be useful in predicting behavior (Cugliandolo and Kurchan 1993; Cugliandolo *et al.* 1997; Berthier *et al.* 2001; Makse and Kurchan 2002; Ono *et al.* 2002; Abou and Gallet 2004; Feitosa and Menon 2004; Ojha *et al.* 2004; Haxton and Liu 2007). Important issues are to elucidate the conditions under which effective temperatures provide a reasonable description, and extent of the analogy to ordinary temperature.

**Formal Theoretical Developments**

One of the great challenges of far-from-equilibrium systems is to develop a theoretical framework, akin to equilibrium and near-equilibrium thermodynamics and statistical mechanics, for tackling these systems. In the last decade, there has been substantial progress in generalizing thermodynamics and statistical mechanics to far-from-equilibrium systems. Steady-state thermodynamics takes into account the heat that is continually generated in steadily-driven systems to generalize the second law of thermodynamics (Oono and Paniconi 1998; Taniguchi and Cohen 2008). Other approaches generalize the concept of entropy to zero-temperature systems (Barrat *et al.* 2001; Blumenfeld and Edwards 2003; Edwards 2004; Henkes *et al.* 2007), while still others generalize the fluctuation-dissipation theorem to far-from-equilibrium systems (Falcioni *et al.* 1990; Ruelle 1999; Crisanti and Ritort 2003; Harada and Sasa 2005; Speck and Seifert 2006). A new thermodynamic result has made it possible to extract equilibrium free energy differences from far-from-equilibrium processes (Jarzynski 1997): the Jarzynski equality states that

$$\exp\left(-\Delta F/k_B T\right) = \left\langle \exp(-W/k_B T\right\rangle \qquad (5)$$

where $\Delta F$ is the free energy difference between two thermodynamic states $A$ and $B$, $k_B$ is the Boltzmann constant, T is the temperature, and W is the work done during a non-equilibrium process that takes the system from A to B. The angular brackets denote an average over all realizations of this process. This equality has been tested experimentally (Liphardt *et al.* 2002; Cohen and Mauzerall 2004) and promises to be an especially useful tool for studying free energy differences in biological systems. Fluctuation theorems have been used to show how irreversibility can emerge from underlying reversible dynamics (Evans and Searles 2002; Sevick *et al.* 2008). Due to these developments, the field of nonequilibrium thermodynamics and statistical mechanics is gathering additional momentum.

**Getting (Un-)Stuck: Jammed States and Jamming Transitions**

The prototypical example of a jammed state is a glass, a state that has both fluid- and solid-like attributes: it has the amorphous structure of a liquid, yet responds to an applied stress like a solid (Ediger *et al.* 1996). All liquids will form glasses upon cooling if crystallization can be avoided (for example, by cooling rapidly enough), and for complex fluids such as polymers, the transition to a glass (plastic) is nearly impossible to avoid. As a liquid is cooled, the time required to reach equilibrium, the relaxation time, increases and the response of the system to perturbations turns more and more sluggish until it becomes immeasurably slow. At this point, the system is now called a glass. The increase of relaxation time is continuous but occurs over an incredibly narrow range of temperature, so that lowering the temperature by 10-20K can increase viscosity and relaxation time by ten or more orders of magnitude. Because the relaxation time exceeds any measurable time scale as the glassy state is approached, a glass by definition is a system far from equilibrium.

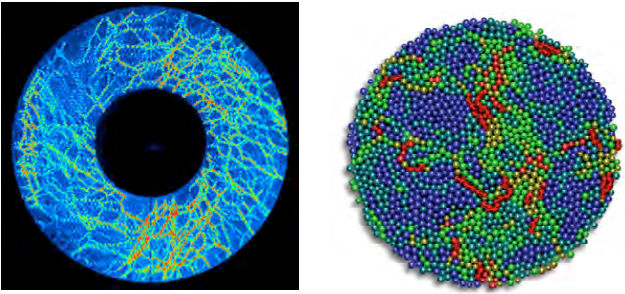

Fig.12. In systems at or near the jamming transition, network-like structures are formed dynamically, such as chains of particles experiencing high contact forces in slowly sheared granular systems (left; (Howell *et al.* 1999)) and strings of particles whose motion is correlated in a gas-fluidized granular bed (right; (Keys *et al.* 2007)) .

Similar glassy states are found not only in ordinary liquids, but in many electronic systems in the presence of disorder, including interacting electron spin systems (spin glasses) or systems of interacting magnetic flux bundles (vortex glasses) (Edwards and Anderson 1975; Blatter *et al.* 1994; Nattermann and Scheidl 2000; Salamon and Jaime 2001). They also occur whenever particles of any size congregate at sufficiently high concentrations (Liu and Nagel 2001)(Fig. 12). For example, micelles or colloids in dense suspensions (Kegel and van Blaaderen 2000; Weeks *et al.* 2000; Trappe *et al.* 2001; Anderson and Lekkerkerker 2002), lubricants trapped between surfaces (Thompson *et al.* 1992; Hu and Granick 1998), bubbles in foams (Durian *et al.* 1991; Sollich *et al.* 1997; Gopal and Durian 1999; Weaire and Hutzler 1999; Banhart 2001), and candies in a jar (Donev *et al.* 2004) all get trapped in glassy states. The onset of glassy behavior is easily observed in an hourglass filled with sand: a fluid-like stream of grains falling through the central neck is rapidly quenched into a solid-like heap that retains the stream's amorphous structure but, unlike a fluid, supports a finite angle of repose. However, once the particles become macroscopic as in the case of sand grains, temperature is no longer effective in facilitating escape from the glassy state. Instead, mechanical fields such as stress or vibration can take over this role and unjam the system (Jaeger *et al.* 1996). The suggestion that temperature and stress can act similarly in systems close to the onset of rigidity has led to the introduction of a more general framework, the concept of jamming (Cates *et al.* 1998; Liu and Nagel 1998; Liu and Nagel 2001; Trappe *et al.* 2001; O'Hern *et al.* 2003). This concept describes the cooperative phenomenon of jamming in terms of the interplay of three key parameters: random thermal motion, applied forcing, and geometrical constraints (Fig. 13).

The idea of a general jamming transition, applying to both thermal and non-thermal systems, has put the spotlight onto some of the most longstanding problems in condensed matter physics, such as the glass transition. Because the jammed state is out of equilibrium, even the most basic questions about any jamming transition remain intensely controversial. Is there a true thermodynamic transition, at which the relaxation time diverges? Or is there a dynamical transition to the jammed state, where the relaxation time diverges with no thermodynamic signature? Or is there no transition at all, so that the relaxation time only truly diverges at zero temperature or mechanical driving? It is because these fundamental questions remain unresolved that the onset of glassy behavior is generally considered one of the most intriguing unsolved problems in condensed matter physics.

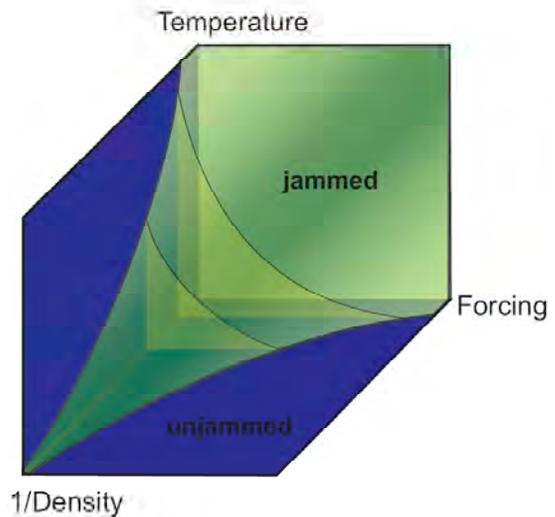

Fig.13. Jamming phase diagram, delineating the conditions under which a multitude of systems become rigid and solid-like. Inside the jammed region (green), these systems are far from equilibrium.

## SUMMARY

Far-from-equilibrium behavior is emerging as one of the major challenges within condensed matter physics and beyond. The importance of making progress in this field is underlined by several key facts: First, far-from-equilibrium behavior is not rare but ubiquitous, occurring from the nanometer scale on up, in daily life as well as high-tech applications. Second, it connects directly to critical, national needs for the next decade, affecting a large fraction of the manufacturing base as well as our economy, climate and environment. Third, we emphasize that far-from-equilibrium behavior cannot be understood simply through small modifications of equilibrium physics. Because it differs so strikingly and at the same time represents largely uncharted intellectual territory, it provides exciting opportunities for major scientific breakthroughs.

The field of far-from-equilibrium physics is vast, and it is unlikely that any one organizing principle will work for all far-from-equilibrium systems. Nonetheless, there is great value in identifying *classes* of systems that might have common underlying physics or that might be tackled by common methods. Recent work within the condensed matter community has set the stage for fresh approaches to longstanding problems concerning far-from-equilibrium behavior by introducing new model systems, such as granular matter, new unifying paradigms, such as jamming, new organizing principles, such as robustness, and new formal approaches, such as steady-state thermodynamics. We are also finding important connections to a wide range of other fields, both within and outside physics. These connections are likely to amplify the impact of advances in far-from-equilibrium physics even further.

Abou, B. and F. Gallet (2004). "Probing a nonequilibrium Einstein relation in an aging colloidal glass." Physical Review Letters **93**(16).

Albert, R. and A. L. Barabasi (2002). "Statistical mechanics of complex networks." Reviews Of Modern Physics **74**(1): 47-97.

Albert, R., H. Jeong and A. L. Barabasi (2000). "Error and attack tolerance of complex networks." Nature **406**(6794): 378-382.

Allen, C., D. Maysinger and A. Eisenberg (1999). "Nano-engineering block copolymer aggregates for drug delivery." Colloids And Surfaces B-Biointerfaces **16**(1-4): 3-27.

Alon, U., M. G. Surette, N. Barkai and S. Leibler (1999). "Robustness in bacterial chemotaxis." Nature **397**(6715): 168-171.

Altshuler, E. and T. H. Johansen (2004). "Colloquium: Experiments in vortex avalanches." Reviews Of Modern Physics **76**(2): 471-487.

Anderson, V. J. and H. N. W. Lekkerkerker (2002). "Insights into phase transition kinetics from colloid science." Nature **416**(6883): 811-815.

Angelescu, D. E., J. H. Waller, D. H. Adamson, P. Deshpande, S. Y. Chou, et al. (2004). "Macroscopic orientation of block copolymer cylinders in single-layer films by shearing." Advanced Materials **16**(19): 1736-+.

Aranson, I. S., A. Gurevich, M. S. Welling, R. J. Wijngaarden, V. K. Vlasko-Vlasov, et al. (2005). "Dendritic flux avalanches and nonlocal electrodynamics in thin superconducting films." Physical Review Letters **94**(3).

Auras, R., B. Harte and S. Selke (2004). "An overview of polylactides as packaging materials." Macromolecular Bioscience **4**(9): 835-864.

Banhart, J. (2001). "Manufacture, characterisation and application of cellular metals and metal foams." Progress In Materials Science **46**(6): 559-U553.

Barkai, N. and S. Leibler (1997). "Robustness in simple biochemical networks." Nature **387**(6636): 913-917.

Barrat, A., J. Kurchan, V. Loreto and M. Sellitto (2001). "Edwards' measures: A thermodynamic construction for dense granular media and glasses." Physical Review E **6305**(5).

Behringer, R. P., K. E. Daniels, T. S. Majmudar and M. Sperl (2008). "Fluctuations, correlations and transitions in granular materials: statistical mechanics for a non-conventional system." Philosophical Transactions Of The Royal Society A-Mathematical Physical And Engineering Sciences **366**(1865): 493-504.

Ben-Jacob, E., I. Cohen and H. Levine (2000). "Cooperative self-organization of microorganisms." Advances In Physics **49**(4): 395-554.

Benzi, R., L. Biferale, R. T. Fisher, L. P. Kadanoff, D. Q. Lamb, et al. (2008). "Intermittency and universality in fully developed inviscid and weakly compressible turbulent flows." Physical Review Letters **100**(23).

Berthier, L., L. F. Cugliandolo and J. L. Iguain (2001). "Glassy systems under time-dependent driving forces: Application to slow granular rheology." Physical Review E **6305**(5).

Bettelheim, E., A. G. Abanov and P. Wiegmann (2006). "Nonlinear quantum shock waves in fractional quantum hall edge states." Physical Review Letters **97**(24).

Bigioni, T. P., X. M. Lin, T. T. Nguyen, E. I. Corwin, T. A. Witten, et al. (2006). "Kinetically driven self assembly of highly ordered nanoparticle monolayers." Nature Materials **5**(4): 265-270.

Blatter, G., M. V. Feigelman, V. B. Geshkenbein, A. I. Larkin and V. M. Vinokur (1994). "Vortices In High-Temperature Superconductors." Reviews Of Modern Physics **66**(4): 1125-1388.

Blumenfeld, R. and S. F. Edwards (2003). "Granular entropy: Explicit calculations for planar assemblies." Physical Review Letters **90**(11).

Bretz, M., R. Zaretzki, S. B. Field, N. Mitarai and F. Nori (2006). "Broad distribution of stick-slip events in slowly sheared granular media: Table-top production of a Gutenberg-Richter-like distribution." Europhysics Letters **74**(6): 1116-1122.

Brujic, J., S. F. Edwards, D. V. Grinev, I. Hopkinson, D. Brujic, et al. (2003). "3D bulk measurements of the force distribution in a compressed emulsion system." Faraday Discussions **123**: 207-220.

Carlson, J. M. and J. Doyle (2002). "Complexity and robustness." Proceedings Of The National Academy Of Sciences Of The United States Of America **99**: 2538-2545.

Carlson, J. M., J. S. Langer and B. E. Shaw (1994). "Dynamics Of Earthquake Faults." Reviews Of Modern Physics **66**(2): 657-670.

Cates, M. E., J. P. Wittmer, J. P. Bouchaud and P. Claudin (1998). "Jamming, force chains, and fragile matter." Physical Review Letters **81**(9): 1841-1844.

Chaikin, P. M. and T. C. Lubensky (1995). Principles of Condensed Matter Physics. Cambridge, UK, Cambridge University Press.

Chen, Z. R., J. A. Kornfield, S. D. Smith, J. T. Grothaus and M. M. Satkowski (1997). "Pathways to macroscale order in nanostructured block copolymers." Science **277**(5330): 1248-1253.

Christie, D. R. (1989). "Long Nonlinear Waves in the Lower Atmosphere." J. Atmospheric Sciences **46**(11): 1462-1491.

Cohen, E. G. D. and D. Mauzerall (2004). "A note on the Jarzynski equality." Journal Of Statistical Mechanics-Theory And Experiment.

Cohen, I., M. P. Brenner, J. Eggers and S. R. Nagel (1999). "Two fluid drop snap-off problem: Experiments and theory." Physical Review Letters **83**(6): 1147-1150.

Cohen, I., H. Li, J. L. Hougland, M. Mrksich and S. R. Nagel (2001). "Using selective withdrawal to coat microparticles." Science **292**(5515): 265-267.

Corwin, E. I., H. M. Jaeger and S. R. Nagel (2005). "Structural signature of jamming in granular media." Nature **435**(7045): 1075-1078.

Crisanti, A. and F. Ritort (2003). "Violation of the fluctuation-dissipation theorem in glassy systems: basic notions and the numerical evidence." Journal Of Physics A-Mathematical And General **36**(21): R181-R290.

Cross, M. C. and P. C. Hohenberg (1993). "Pattern-Formation Outside Of Equilibrium." Reviews Of Modern Physics **65**(3): 851-1112.

Cugliandolo, L. F. and J. Kurchan (1993). "Analytical Solution Of The Off-Equilibrium Dynamics Of A Long-Range Spin-Glass Model." Physical Review Letters **71**(1): 173-176.

Cugliandolo, L. F., J. Kurchan and L. Peliti (1997). "Energy flow, partial equilibration, and effective temperatures in systems with slow dynamics." Physical Review E **55**(4): 3898-3914.

de Gennes, P. G. (1999). "Granular matter: a tentative view." Reviews Of Modern Physics **71**(2): S374-S382.

Debenedetti, P. G. and F. H. Stillinger (2001). "Supercooled liquids and the glass transition." Nature **410**(6825): 259-267.

DiDonna, B. A. and T. A. Witten (2001). "Anomalous strength of membranes with elastic ridges." Physical Review Letters **8720**(20).

Donev, A., I. Cisse, D. Sachs, E. Variano, F. H. Stillinger, et al. (2004). "Improving the density of jammed disordered packings using ellipsoids." Science **303**(5660): 990-993.

Doshi, P., I. Cohen, W. W. Zhang, M. Siegel, P. Howell, et al. (2003). "Persistence of memory in drop breakup: The breakdown of universality." Science **302**(5648): 1185-1188.

Durian, D. J., D. A. Weitz and D. J. Pine (1991). "Scaling Behavior in Shaving Cream." Physical Review A **44**(12): R7902-R7905.

Ediger, M. D., C. A. Angell and S. R. Nagel (1996). "Supercooled liquids and glasses." Journal Of Physical Chemistry **100**(31): 13200-13212.

Edwards, S. F. (2004). "New kinds of entropy." Journal Of Statistical Physics **116**(1-4): 29-42.

Edwards, S. F. and P. W. Anderson (1975). "Theory Of Spin Glasses." Journal Of Physics F-Metal Physics **5**(5): 965-974.

Eggers, J. and E. Villermaux (2008). "Physics of liquid jets." Reports On Progress In Physics **71**(3).

Ennis, B. J., J. Green and R. Davies (1994). "The Legacy Of Neglect In The United-States." Chemical Engineering Progress **90**(4): 32-43.

Evans, D. J. and D. J. Searles (2002). "The fluctuation theorem." Advances In Physics **51**(7): 1529-1585.

Faisst, H. and B. Eckhardt (2004). "Sensitive dependence on initial conditions in transition to turbulence in pipe flow." Journal Of Fluid Mechanics **504**: 343-352.

Falcioni, M., S. Isola and A. Vulpiani (1990). "Correlation-Functions And Relaxation Properties In Chaotic Dynamics And Statistical-Mechanics." Physics Letters A **144**(6-7): 341-346.

Falkovich, G., K. Gawedzki and M. Vergassola (2001). "Particles and fields in fluid turbulence." Reviews Of Modern Physics **73**(4): 913-975.

Feitosa, K. and N. Menon (2004). "Fluidized granular medium as an instance of the fluctuation theorem." Physical Review Letters **92**(16).

Field, S., J. Witt, F. Nori and X. S. Ling (1995). "Superconducting Vortex Avalanches." Physical Review Letters **74**(7): 1206-1209.

Fisker, J. L., D. S. Balsara and T. Burger (2006). "The accretion and spreading of matter on white dwarfs." New Astronomy Reviews **50**(7-8): 509-515.

Freund, L. B. (1998). Dynamic Fracture Mechanics. Cambridge, Cambridge University Press.

Gao, H. J., B. H. Ji, I. L. Jager, E. Arzt and P. Fratzl (2003). "Materials become insensitive to flaws at nanoscale: Lessons from nature." Proceedings Of The National Academy Of Sciences Of The United States Of America **100**(10): 5597-5600.

Gaspard, P. (2006). "Out-of-equilibrium nanosystems." Progress Of Theoretical Physics Supplement(165): 33-56.

Ghashghaie, S., W. Breymann, J. Peinke, P. Talkner and Y. Dodge (1996). "Turbulent cascades in foreign exchange markets." Nature **381**(6585): 767-770.

Gopal, A. D. and D. J. Durian (1999). "Shear-induced "melting" of an aqueous foam." Journal of Colloid and Interface Science **213**(1): 169-178.

Groisman, A. and V. Steinberg (2000). "Elastic turbulence in a polymer solution flow." Nature **405**(6782): 53-55.

Guinotte, J. M. and V. J. Fabry (2008). Ocean acidification and its potential effects on marine ecosystems. Year In Ecology And Conservation Biology 2008. **1134:** 320-342.

Gunes, I. S. and S. C. Jana (2008). "Shape memory polymers and their nanocomposites: A review of science and technology of new multifunctional materials." Journal Of Nanoscience And Nanotechnology **8**(4): 1616-1637.

Harada, T. and S. Sasa (2005). "Equality connecting energy dissipation with a violation of the fluctuation-response relation." Physical Review Letters **95**(13).

Hatano, M., S. Moon, M. Lee, K. Suzuki and C. P. Grigoropoulos (2000). "Excimer laser-induced temperature field in melting and resolidification of silicon thin films." Journal Of Applied Physics **87**(1): 36-43.

Hawker, C. J. and T. P. Russell (2005). "Block copolymer lithography: Merging "bottom-up" with "top-down" processes." Mrs Bulletin **30**(12): 952-966.

Haxton, T. K. and A. J. Liu (2007). "Activated dynamics and effective temperature in a steady state sheared glass." Physical Review Letters **99**(19).

Helbing, D. (2001). "Traffic and related self-driven many-particle systems." Reviews Of Modern Physics **73**(4): 1067-1141.

Henkes, S., C. S. O'Hern and B. Chakraborty (2007). "Entropy and temperature of a static granular assembly: An ab initio approach." Physical Review Letters **99**(3).

Hof, B., A. Juel and T. Mullin (2003). "Scaling of the turbulence transition threshold in a pipe." Physical Review Letters **91**(24).

Howell, D., R. P. Behringer and C. Veje (1999). "Stress fluctuations in a 2D granular Couette experiment: A continuous transition." Physical Review Letters **82**(26): 5241-5244.

Hu, Y. Z. and S. Granick (1998). "Microscopic study of thin film lubrication and its contributions to macroscopic tribology." Tribology Letters **5**(1): 81-88.

Huang, Z. F., F. Chen, R. D'Agosta, P. A. Bennett, M. Di Ventra, et al. (2007). "Local ionic and electron heating in single-molecule junctions." Nature Nanotechnology **2**(11): 698-703.

Inoue, A. (2000). "Stabilization of metallic supercooled liquid and bulk amorphous alloys." Acta Materialia **48**(1): 279-306.

Jaeger, H. M. and S. R. Nagel (1992). "Physics Of The Granular State." Science **255**(5051): 1523-1531.

Jaeger, H. M., S. R. Nagel and R. P. Behringer (1996). "Granular solids, liquids, and gases." Reviews Of Modern Physics **68**(4): 1259-1273.

Jarzynski, C. (1997). "Nonequilibrium equality for free energy differences." Physical Review Letters **78**(14): 2690-2693.

Jerne, N. K. (1974). "Towards a network theory of immune system." Ann. Immunol. **C125**: 373-389.

Kadanoff, L., D. Lohse, J. Wang and R. Benzi (1995). "Scaling And Dissipation In The Goy Shell-Model." Physics Of Fluids **7**(3): 617-629.

Kadanoff, L. P. (1999). "Built upon sand: Theoretical ideas inspired by granular flows." Reviews Of Modern Physics **71**(1): 435-444.

Kegel, W. K. and A. van Blaaderen (2000). "Direct observation of dynamical heterogeneities in colloidal hard-sphere suspensions." Science **287**(5451): 290-293.

Keim, N. C., P. Moller, W. W. Zhang and S. R. Nagel (2006). "Breakup of air bubbles in water: Memory and breakdown of cylindrical symmetry." Physical Review Letters **97**(14).

Keys, A. S., A. R. Abate, S. C. Glotzer and D. J. Durian (2007). "Measurement of growing dynamical length scales and prediction of the jamming transition in a granular material." Nature Physics **3**(4): 260-264.

Kim, S. O., H. H. Solak, M. P. Stoykovich, N. J. Ferrier, J. J. de Pablo, et al. (2003). "Epitaxial self-assembly of block copolymers on lithographically defined nanopatterned substrates." Nature **424**(6947): 411-414.

Knowlton, T. M., J. W. Carson, G. E. Klinzing and W. C. Yang (1994). "The Importance Of Storage, Transfer, And Collection." Chemical Engineering Progress **90**(4): 44-54.

La Porta, A., G. A. Voth, A. M. Crawford, J. Alexander and E. Bodenschatz (2001). "Fluid particle accelerations in fully developed turbulence." <u>Nature</u> **409**(6823): 1017-1019.

Langer, J. S. (1980). "Instabilities And Pattern-Formation In Crystal-Growth." <u>Reviews Of Modern Physics</u> **52**(1): 1-28.

Lee, D., S. G. Jia, S. Banerjee, J. Bevk, I. P. Herman, et al. (2007). "Viscoplastic and granular behavior in films of colloidal nanocrystals." <u>Physical Review Letters</u> **98**(2).

Lewis, G. S. and H. L. Swinney (1999). "Velocity structure functions, scaling, and transitions in high-Reynolds-number Couette-Taylor flow." <u>Physical Review E</u> **59**(5): 5457-5467.

Lin, P. Z. and P. L. F. Liu (1998). "A numerical study of breaking waves in the surf zone." <u>Journal Of Fluid Mechanics</u> **359**: 239-264.

Lin, Y., H. Skaff, A. Boker, A. D. Dinsmore, T. Emrick, et al. (2003). "Ultrathin cross-linked nanoparticle membranes." <u>Journal Of The American Chemical Society</u> **125**(42): 12690-12691.

Liphardt, J., S. Dumont, S. B. Smith, I. Tinoco and C. Bustamante (2002). "Equilibrium information from nonequilibrium measurements in an experimental test of Jarzynski's equality." <u>Science</u> **296**(5574): 1832-1835.

Liu, A. J. and S. R. Nagel (1998). "Nonlinear dynamics - Jamming is not just cool any more." <u>Nature</u> **396**(6706): 21-22.

Liu, A. J. and S. R. Nagel (2001). <u>Jamming and rheology : constrained dynamics on microscopic and macroscopic scales</u>. London, Taylor & Francis.

Liu, C. H., S. R. Nagel, D. A. Schecter, S. N. Coppersmith, S. Majumdar, et al. (1995). "Force Fluctuations In Bead Packs." <u>Science</u> **269**(5223): 513-515.

Lopes, W. A. and H. M. Jaeger (2001). "Hierarchical self-assembly of metal nanostructures on diblock copolymer scaffolds." <u>Nature</u> **414**(6865): 735-738.

Low, B. C. (2001). "Coronal mass ejections, magnetic flux ropes, and solar magnetism." <u>Journal Of Geophysical Research-Space Physics</u> **106**(A11): 25141-25163.

Majmudar, T. S. and R. P. Behringer (2005). "Contact force measurements and stress-induced anisotropy in granular materials." <u>Nature</u> **435**(7045): 1079-1082.

Makse, H. A. and J. Kurchan (2002). "Testing the thermodynamic approach to granular matter with a numerical model of a decisive experiment." <u>Nature</u> **415**(6872): 614-617.

Mantegna, R. N. and H. E. Stanley (2000). <u>An introduction to econophysics: correlations and complexity in finance</u>. New York, Cambridge University Press.

Marcus, P. S. (1993). "Jupiter Great Red Spot And Other Vortices." <u>Annual Review Of Astronomy And Astrophysics</u> **31**: 523-573.

Marder, M. and J. Fineberg (1996). "How things break." <u>Physics Today</u> **49**(9): 24-29.

McCauley, J. L. (2004). <u>Dynamics of Markets: Econophysics and Finance</u>. Cambridge, UK, Cambridge University Press.

Mellor, G. L. and T. Yamada (1974). "Hierarchy Of Turbulence Closure Models For Planetary Boundary-Layers." <u>Journal Of The Atmospheric Sciences</u> **31**(7): 1791-1806.

Merrow, E. W. (1985). "Linking R&D To Problems Experienced In Solids Processing." <u>Chemical Engineering Progress</u> **81**(5): 14-22.

Merrow, E. W. (1988). "Estimating start-up times for solids processing Plants." <u>Chemical Engineering</u> **95**(15): 89-92.

Meulenbroek, B., C. Storm, V. Bertola, C. Wagner, D. Bonn, et al. (2003). "Intrinsic route to melt fracture in polymer extrusion: A weakly nonlinear subcritical instability of viscoelastic poiseuille flow." <u>Physical Review Letters</u> **90**(2).

Milo, R., S. Shen-Orr, S. Itzkovitz, N. Kashtan, D. Chklovskii, et al. (2002). "Network motifs: Simple building blocks of complex networks." Science **298**(5594): 824-827.

Moritz, M. A., M. E. Morais, L. A. Summerell, J. M. Carlson and J. Doyle (2005). "Wildfires, complexity and highly optimized tolerance." Proceedings Of The National Academy Of Sciences Of The United States Of America **102**: 17912-17917.

Morozov, A. N. and W. van Saarloos (2007). "An introductory essay on subcritical instabilities and the transition to turbulence in visco-elastic parallel shear flows." Physics Reports-Review Section Of Physics Letters **447**(3-6): 112-143.

Murty, B. S. and S. Ranganathan (1998). "Novel materials synthesis by mechanical alloying/milling." International Materials Reviews **43**(3): 101-141.

Nasuno, S., A. Kudrolli and J. P. Gollub (1997). "Friction in granular layers: Hysteresis and precursors." Physical Review Letters **79**(5): 949-952.

Nattermann, T. and S. Scheidl (2000). "Vortex-glass phases in type-II superconductors." Advances In Physics **49**(5): 607-704.

Newman, M. E. J. (2003). "The structure and function of complex networks." Siam Review **45**(2): 167-256.

Nowak, E. R., O. W. Taylor, L. Liu, H. M. Jaeger and T. I. Selinder (1997). "Magnetic flux instabilities in superconducting niobium rings: Tuning the avalanche behavior." Physical Review B **55**(17): 11702-11705.

NRC, BPA and Committee on CMMP 2010 (2007). Condensed Matter And Materials Physics: The Science of the World Around Us. Washington, D.C., The National Academies Press.

O'Hern, C. S., L. E. Silbert, A. J. Liu and S. R. Nagel (2003). "Jamming at zero temperature and zero applied stress: The epitome of disorder." Physical Review E **68**(1).

Ogawa, M. (2008). "Mantle convection: A review." Fluid Dynamics Research **40**(6): 379-398.

Ojha, R. P., P. A. Lemieux, P. K. Dixon, A. J. Liu and D. J. Durian (2004). "Statistical mechanics of a gas-fluidized particle." Nature **427**(6974): 521-523.

Olson, C. J., C. Reichhardt and F. Nori (1997). "Superconducting vortex avalanches, voltage bursts, and vortex plastic flow: Effect of the microscopic pinning landscape on the macroscopic properties." Physical Review B **56**(10): 6175-6194.

Ono, I. K., C. S. O'Hern, D. J. Durian, S. A. Langer, A. J. Liu, et al. (2002). "Effective temperatures of a driven system near jamming." Physical Review Letters **89**(9).

Oono, Y. and M. Paniconi (1998). "Steady state thermodynamics." Progress Of Theoretical Physics Supplement(130): 29-44.

Orr, J. C., V. J. Fabry, O. Aumont, L. Bopp, S. C. Doney, et al. (2005). "Anthropogenic ocean acidification over the twenty-first century and its impact on calcifying organisms." Nature **437**(7059): 681-686.

Ott, S. and J. Mann (2000). "An experimental investigation of the relative diffusion of particle pairs in three-dimensional turbulent flow." Journal Of Fluid Mechanics **422**: 207-223.

Ouellette, N. T., H. T. Xu and E. Bodenschatz (2006). "A quantitative study of three-dimensional Lagrangian particle tracking algorithms." Experiments In Fluids **40**(2): 301-313.

Palmer, T. N. (1993). "Extended-Range Atmospheric Prediction And The Lorenz Model." Bulletin Of The American Meteorological Society **74**(1): 49-65.

Park, C., J. Yoon and E. L. Thomas (2003). "Enabling nanotechnology with self assembled block copolymer patterns." Polymer **44**(22): 6725-6760.

Park, M., C. Harrison, P. M. Chaikin, R. A. Register and D. H. Adamson (1997). "Block copolymer lithography: Periodic arrays of similar to 10(11) holes in 1 square centimeter." Science **276**(5317): 1401-1404.

Plewa, T., A. C. Calder and D. Q. Lamb (2004). "Type Ia supernova explosion: Gravitationally confined detonation." Astrophysical Journal **612**(1): L37-L40.

Proulx, S. R., D. E. L. Promislow and P. C. Phillips (2005). "Network thinking in ecology and evolution." Trends in Ecology and Evolution **20**: 343-353.

Rabani, E., D. R. Reichman, P. L. Geissler and L. E. Brus (2003). "Drying-mediated self-assembly of nanoparticles." Nature **426**(6964): 271-274.

Roberts, P. H. and G. A. Glatzmaier (2000). "Geodynamo theory and simulations." Reviews Of Modern Physics **72**(4): 1081-1123.

Ross, C. A., H. I. Smith, T. Savas, M. Schattenburg, M. Farhoud, et al. (1999). "Fabrication of patterned media for high density magnetic storage." Journal Of Vacuum Science & Technology B **17**(6): 3168-3176.

Ruelle, D. (1999). "Smooth dynamics and new theoretical ideas in nonequilibrium statistical mechanics." Journal Of Statistical Physics **95**(1-2): 393-468.

Salamon, M. B. and M. Jaime (2001). "The physics of manganites: Structure and transport." Reviews Of Modern Physics **73**(3): 583-628.

Segre, P. N., F. Liu, P. Umbanhowar and D. A. Weitz (2001). "An effective gravitational temperature for sedimentation." Nature **409**(6820): 594-597.

Sevick, E. M., R. Prabhakar, S. R. Williams and D. J. Searles (2008). "Fluctuation theorems." Annual Review Of Physical Chemistry **59**: 603-633.

Shinbrot, T. and F. J. Muzzio (2000). "Nonequilibrium patterns in granular mixing and segregation." Physics Today **53**(3): 25-30.

Siggia, E. D. (1994). "High Rayleigh Number Convection." Annual Review Of Fluid Mechanics **26**: 137-168.

Sollich, P., F. Lequeux, P. Hebraud and M. E. Cates (1997). "Rheology of soft glassy materials." Physical Review Letters **78**(10): 2020-2023.

Sornette, D. (2003). Why Stock Markets Crash: Critical Events in Complex Financial Systems. Princeton, NY, Princeton University Press.

Speck, T. and U. Seifert (2006). "Restoring a fluctuation-dissipation theorem in a nonequilibrium steady state." Europhysics Letters **74**(3): 391-396.

Sposili, R. S. and J. S. Im (1996). "Sequential lateral solidification of thin silicon films on SiO2." Applied Physics Letters **69**(19): 2864-2866.

Sreenivasan, K. R. (1999). "Fluid turbulence." Reviews Of Modern Physics **71**: S383-S395.

Stoldt, C. R., A. M. Cadilhe, C. J. Jenks, J. M. Wen, J. W. Evans, et al. (1998). "Evolution of far-from-equilibrium nanostructures formed by cluster-step and cluster-cluster coalescence in metal films." Physical Review Letters **81**(14): 2950-2953.

Strogatz, S. H. (2001). "Exploring complex networks." Nature **410**(6825): 268-276.

Suryanarayana, C. (2001). "Mechanical alloying and milling." Progress In Materials Science **46**(1-2): 1-184.

Sztrum, C. G. and E. Rabani (2006). "Out-of-equilibrium self-assembly of binary mixtures of nanoparticles." Advanced Materials **18**(5): 565-+.

Tabeling, P. (2002). "Two-dimensional turbulence: a physicist approach." Physics Reports-Review Section Of Physics Letters **362**: 1-62.

Taniguchi, T. and E. G. D. Cohen (2008). "Nonequilibrium steady state thermodynamics and fluctuations for stochastic systems." Journal Of Statistical Physics **130**(4): 633-667.

Terris, B. D. and T. Thomson (2005). "Nanofabricated and self-assembled magnetic structures as data storage media." Journal Of Physics D-Applied Physics **38**(12): R199-R222.

Thompson, P. A., G. S. Grest and M. O. Robbins (1992). "Phase-Transitions And Universal Dynamics In Confined Films." Physical Review Letters **68**(23): 3448-3451.

Thompson, R. B., V. V. Ginzburg, M. W. Matsen and A. C. Balazs (2001). "Predicting the mesophases of copolymer-nanoparticle composites." Science **292**(5526): 2469-2472.

Toner, J., Y. H. Tu and S. Ramaswamy (2005). "Hydrodynamics and phases of flocks." Annals Of Physics **318**(1): 170-244.

Trappe, V., V. Prasad, L. Cipelletti, P. N. Segre and D. A. Weitz (2001). "Jamming phase diagram for attractive particles." Nature **411**(6839): 772-775.

Tyson, J. J., R. Albert, A. Goldbeter, P. Ruoff and J. Sible (2008). "Biological switches and clocks." Journal of the Royal Society Interface **5**: S1-S8.

Vandenberghe, N., J. Zhang and S. Childress (2004). "Symmetry breaking leads to forward flapping flight." Journal Of Fluid Mechanics **506**: 147-155.

Wang, W. H., C. Dong and C. H. Shek (2004). "Bulk metallic glasses." Materials Science & Engineering R-Reports **44**(2-3): 45-89.

Weaire, D. and S. Hutzler (1999). The Physics of Foams. Oxford, UK, Oxford University Press.

Weeks, E. R., J. C. Crocker, A. C. Levitt, A. Schofield and D. A. Weitz (2000). "Three-dimensional direct imaging of structural relaxation near the colloidal glass transition." Science **287**(5453): 627-631.

White, C. M. and M. G. Mungal (2008). "Mechanics and prediction of turbulent drag reduction with polymer additives." Annual Review Of Fluid Mechanics **40**: 235-256.

Witten, T. A. (2007). "Stress focusing in elastic sheets." Reviews Of Modern Physics **79**(2): 643-675.

Wu, J. G. and O. L. Loucks (1995). "From balance of nature to hierarchical patch dynamics: A paradigm shift in ecology." Quarterly Review Of Biology **70**(4): 439-466.

Wuttig, M. and N. Yamada (2007). "Phase-change materials for rewriteable data storage." Nature Materials **6**(11): 824-832.

Wyman, J. L., S. Kizilel, R. Skarbek, X. Zhao, M. Connors, et al. (2007). "Immunoisolating pancreatic islets by encapsulation with selective withdrawal." Small **3**(4): 683-690.

Xu, J., J. F. Xia and Z. Q. Lin (2007). "Evaporation-induced self-assembly of nanoparticles from a sphere-on-flat geometry." Angewandte Chemie-International Edition **46**(11): 1860-1863.

Zakery, A. and S. R. Elliott (2003). "Optical properties and applications of chalcogenide glasses: a review." Journal Of Non-Crystalline Solids **330**(1-3): 1-12.

Zhang, J., S. Childress, A. Libchaber and M. Shelley (2000). "Flexible filaments in a flowing soap film as a model for one-dimensional flags in a two-dimensional wind." Nature **408**(6814): 835-839.

**Image Credits for Figures**

**Figure 1**
Top: school of fish, http://imagecache2.allposters.com/images/PTGPOD/603809.jpg
Middle: hurricane Francis, http://www.shnorth.com/Hurricane.htm
Bottom: spiral galaxy NGC4414, http://en.wikipedia.org/wiki/File:NGC_4414_(NASA-med).jpg

**Figure 2**
Top: wing tip turbulence produced by small agricultural plane,
    http://en.wikipedia.org/wiki/File:Airplane_vortex_edit.jpg
Bottom: Aloha Airlines cabin top peeled off during flight, from Michale Marder's web page on "How things break",http://chaos.ph.utexas.edu/~marder/fracture/phystoday/how_things_break/how_things_break.html

**Figure 3**
Snow flake crystal, http://www.its.caltech.edu/~atomic/snowcrystals

**Figure 4**
Clockwise from top:
molten glass, http://srnl.doe.gov/images/pour.jpg
molten glass, http://www.usefilm.com/image/1163585.html
Styrofoam, http://www.aic.info.ro/styrofoam/
soapy foam, image courtesy of Doug Durian, University of Pennsylvania
simulation of Ni foam, http://www.itwm.fhg.de/mab/projects/MIKRO/nifoam.jpg

**Figure 5**
Left: sand dune, http://www.photo.net/photo/pcd0738/great-sand-dune-ridge-7
Right: http://jfi.uchicago.edu/~jaeger/group/granular.html, image adapted from Fig. 1 of Matthias E. Möbius et al., "Intruders in the Dust: Air-Driven Granular Size Separation", Phys. Rev. Lett., **93**, 198001 (2004).

**Figure 6**
Strength versus elastic limit of various materials, William L. Johnson,
http://www.its.caltech.edu/~matsci/wlj/wlj_research.html

Figure 7
flapping filament flow visualizations, Jun Zhang, NYU,
    http://physics.nyu.edu/~jz11/filaments.html

**Figure 8**
Right: fracture in glass with nickel-sulphide inclusions, http://www.picams.com.au/nickel-sulphide.html
Left: turbulence in clouds over Canary Islands (NASA),
    http://plus.maths.org/issue26/news/turbulence/index.html

**Figure 9**

Left: crumpled sheet with sharp points and ridges,  taken from Tom Witten's talk given at Frontiers of Condensed Matter workshop, Snowmass Aspen, 2004, http://frontiers.physics.rutgers.edu

Right: entrainment of a small object due to (near-)singular flow at an oil/water interface, image courtesy Sid Nagel, see Jason Wyman et al., "Immunoisolating Pancreatic Islets by Encapsulation with Selective Withdrawal," Small **3**, 683–690 (2007).

**Figure 10**

Top: map of interacting yeast proteins, from A.-L. Barabasi and E. Bonabeau, "Scale-Free Networks," Scientific American **288** (5), 50-59 (2003), http://www.nd.edu/~networks/PublicationCategories/ReviewArticles/ScaleFree_ScientificAmeri288,60-69(2003).pdf

Bottom: Internet map,  Date: Nov 22 2003, http://www.opte.org/maps/

**Figure 11**

Left: snow avalanche,  http://www.avalanche.org/pictures/avalanche1.jpg

Middle: granular avalanche in pile of mustard seeds, cover image accompanying H. M. Jaeger, S. R. Nagel, and R. P. Behringer, "The Physics of Granular Materials", Physics Today **49**, 32 (1996).

Right: flux bundle avalanche  in type-II superconductor, Cynthia Olson-Reichhardt, http://www.t12.lanl.gov/home/olson/ and http://www.t12.lanl.gov/home/olson/aval_index.html